\documentclass[preprint,showpacs,preprintnumbers,amsmath,amssymb]{revtex4}

\usepackage{graphicx}% Include figure files
\usepackage{dcolumn}% Align table columns on decimal point
\usepackage{bm}% bold math

\begin{document}

\preprint{APS/123-QED}

\title{ Influence of charge asymmetry and isospin dependent cross-section on elliptical flow\\}%Force line breaks with\\

\author{Anupriya Jain}
\author{Suneel Kumar}%
 \email{suneel.kumar@thapar.edu}
\affiliation{%
School of Physics and Material Science, Thapar University, Patiala-147004, Punjab (India)\\
}%

\author{Rajeev K. Puri}
\affiliation{
Department of Physics, Panjab University, Chandigarh-160014, India\\
% with \\
}%
\date{\today}% It is always \today, today,

\begin{abstract}
Using the isospin dependent quantum molecular dynamics model, we study the effect of charge asymmetry and isospin dependent cross-section on different aspects of elliptical flow. Simulations have been carried out for the reactions of $^{124}X_{m}+^{124}X_{m}$, where m = (47, 50, 53, 57 and 59) and $^{40}X_{n}+^{40}X_{n}$, where n= (14, 16, 18, 21 and 23). Our study shows that elliptical flow depend strongly on the isospin of cross-section. The transition energy remains almost constant with increase in N/Z of the system. A good agreement is obtained with experimental measurements.  
\end{abstract}

\maketitle
\baselineskip=1.5\baselineskip\
\section{Introduction}

Heavy-ion collisions provide a possibility to study the properties of nuclear matter in conditions that are vastly different than reported in normal nuclei \cite{1}. Considerable progress has been made recently in determining the equation of state of nuclear matter from heavy ion reaction data \cite{2}. A prominent role among available observables is played by the collective flow. Much theoretical and experimental efforts have been made to the study of collective flow in HICs \cite{3}. The elliptic flow has proven to be one of the most fruitful probes for extracting the EoS and the dynamics of heavy-ion collisions. The parameter of elliptic flow is quantified by the second-order Fourier coefficient \cite{4} from the azimuthal distribution of detected particles at midrapidity as:

\begin{equation}
\frac{dN}{d\phi} = p_0(1 + 2v_1Cos\phi + 2v_2Cos2\phi).
\end{equation}

Where $\phi$ is the azimuthal angle of emitted particle momentum relative to the x axis. Positive values for $\langle$ $Cos 2\phi$ $\rangle$ reflect a preferential in-plane emission, whereas negative values of $\langle$ $Cos 2\phi$ $\rangle$ reflect a preferential out-of-plane emission.

Yu-Ming Zheng {\it et al.,} \cite{5} studied the proton elliptic flow in the collisions of $^{48}Ca + ^{48}Ca$ at energies from 30 to 100 MeV/nucleon.  They showed that, with increasing incident energy, the elliptic flow shows a transition from positive to negative values. Its magnitude was found to depend on both nuclear equation of state (EOS) as well as on the nucleon-nucleon scattering cross-section. On the other hand, D. Persram {\it et al.,} \cite{6} have shown, for the first time, the effect of parameter-free self-consistent calculation of nucleon-nucleon in-medium scattering cross-section implemented BUU model on different flows. Their results favored in-medium cross-section compared to free one. Y. Zhang {\it et al.,} \cite{7} have investigated the elliptic flow in heavy-ion collisions at energies from several tens to several hundreds of MeV/nucleon. They showed that, soft nuclear equation of state and incident energy dependent in-medium nucleon-nucleon cross-sections are required to describe the excitation function of the elliptic flow at intermediate energies.\\
%%%%%%%%%%%%%%%%%%%%%%%%%%%%%%%%%%%%%%%%%%%%%%%%%%%%%%%%%%%%%%%%%%%%%%%%%
\section{ISOSPIN-DEPENDENT QUANTUM MOLECULAR DYNAMICS (IQMD) MODEL}
Our study is performed within the framework of IQMD \cite{10} model
where hadrons propagate with Hamilton equations of motion:
\begin{equation}
\frac{d{r_i}}{dt}~=~\frac{d\it{\langle~H~\rangle}}{d{p_i}}~~;~~\frac{d{p_i}}{dt}~=~-\frac{d\it{\langle~H~\rangle}}{d{r_i}},
\end{equation}
with
\begin{eqnarray}
\langle~H~\rangle&=&\langle~T~\rangle+\langle~V~\rangle\nonumber\\
&=&\sum_{i}\frac{p_i^2}{2m_i}+
\sum_i \sum_{j > i}\int f_{i}(\vec{r},\vec{p},t)V^{\it ij}({\vec{r}^\prime,\vec{r}})\nonumber\\
& &\times f_j(\vec{r}^\prime,\vec{p}^\prime,t)d\vec{r}d\vec{r}^\prime d\vec{p}d\vec{p}^\prime .
\end{eqnarray}
 The baryon-baryon potential $V^{ij}$, in the above relation, reads as:
\begin{eqnarray}
V^{ij}(\vec{r}^\prime -\vec{r})&=&V^{ij}_{Skyrme}+V^{ij}_{Yukawa}+V^{ij}_{Coul}+V^{ij}_{sym}\nonumber\\
&=& \left [t_{1}\delta(\vec{r}^\prime -\vec{r})+t_{2}\delta(\vec{r}^\prime -\vec{r})\rho^{\gamma-1}
\left(\frac{\vec{r}^\prime +\vec{r}}{2}\right) \right]\nonumber\\
& & +~t_{3}\frac{exp(|\vec{r}^\prime-\vec{r}|/\mu)}{(|\vec{r}^\prime-\vec{r}|/\mu)}~+~\frac{Z_{i}Z_{j}e^{2}}{|\vec{r}^\prime -\vec{r}|}\nonumber\\
& & + t_{6}\frac{1}{\varrho_0}T_3^{i}T_3^{j}\delta(\vec{r_i}^\prime -\vec{r_j}).
\label{s1}
\end{eqnarray}
Here $Z_i$ and $Z_j$ denote the charges of $i^{th}$ and $j^{th}$ baryon, and $T_3^i$, $T_3^j$ are their respective $T_3$
components (i.e. 1/2 for protons and -1/2 for neutrons). Meson potential consists of Coulomb interaction only.
The binary nucleon-nucleon collisions are included by employing the collision term of well known VUU-BUU equation.
 During the propagation, two nucleons are
supposed to suffer a binary collision if the distance between their centroids
\begin{equation}
|r_i-r_j| \le \sqrt{\frac{\sigma_{tot}}{\pi}}, \sigma_{tot} = \sigma(\sqrt{s}, type),
\end{equation}
"type" denotes the ingoing collision partners (N-N, N-$\Delta$, N-$\pi$,..). In addition,
Pauli blocking (of the final
state) of baryons is taken into account by checking the phase space densities in the final states.\\
%%%%%%%%%%%%%%%%%%%%%%%%%%%%%%%%%%%%%%%%%%%%%%%%%%%%%%%%%%%%%%%%%%%%%%%%%
\section{Results and Discussion}
For the present analysis, simulations are carried out for two sets of the reaction using soft equation of state. For the first case, the mass of the colliding nuclei is fixed to be 124 units and for the second set, the  mass of the colliding nuclei is fixed to be 40 units. In other words, we study, the reactions of $^{124}X_{m}+^{124}X_{m}$, where $^{124}X_{m}$ = ($^{124}Ag_{47}$, $^{124}Sn_{50}$, $^{124}I_{53}$, $^{124}La_{57}$ and $^{124}Pr_{59}$), respectively. The second set corresponds to: $^{40}Y_{n}+^{40}Y_{n}$, where $^{40}Y_{n}$ = ($^{40}V_{23}$, $^{40}Sc_{21}$, $^{40}Ar_{18}$, $^{40}S_{16}$ and $^{40}Si_{14}$), respectively. 
The phase space generated by the IQMD model has been analyzed using the minimum spanning tree (MST) \cite{10} method. 
The elliptical flow is defined as the average difference between the square of x and y components of the particles transverse momentum. Mathematically, it can be written as:

\begin{equation}
\langle v_2 \rangle = <Cos2\phi> = \langle \frac{p_x^2 - p_y^2}{p_x^2 + p_y^2}\rangle
\end{equation}

where $p_{x}$ and $p_{y}$ are the x and y components of the momentum. The positive value of elliptical flow describes the eccentricity of an ellipse-like distribution and indicates in-plane enhancement of the particle emission. On the other hand, a negative value of $v_{2}$ shows the squeeze-out effects perpendicular to the reaction plane. Obviously, zero value corresponds to an isotropic distribution. Generally, for a meaningful understanding $\langle$ $v_{2}$ $\rangle$  is extracted from the mid-rapidity region only. Naturally, mid-rapidity region (-0.1 $\leq$ $\frac {Y_{c.m}}{Y_{beam}}$ $\leq$ 0.1) corresponds to the collision (participant) zone and hence signifies compressed matter. On the other hand, $\frac {Y_{c.m}}{Y_{beam}}$ $\neq$ 0 corresponds to the spectator zone, ($\frac {Y_{c.m}}{Y_{beam}}$ $\langle$ -0.1) corresponds to target like (TL) matter and  ($\frac {Y_{c.m}}{Y_{beam}}$ $\rangle$ 0.1) corresponds to projectile like (PL) matter.
%%%%%%%%%%%%%%%%%%%%%%%%%%%%%%%%%%%%%%%%%%%%%%%%%%%%%%%%%%%%%%%%%%%%%%%%%
\begin{figure}
\hspace{-2.0cm}\includegraphics[scale=0.45]{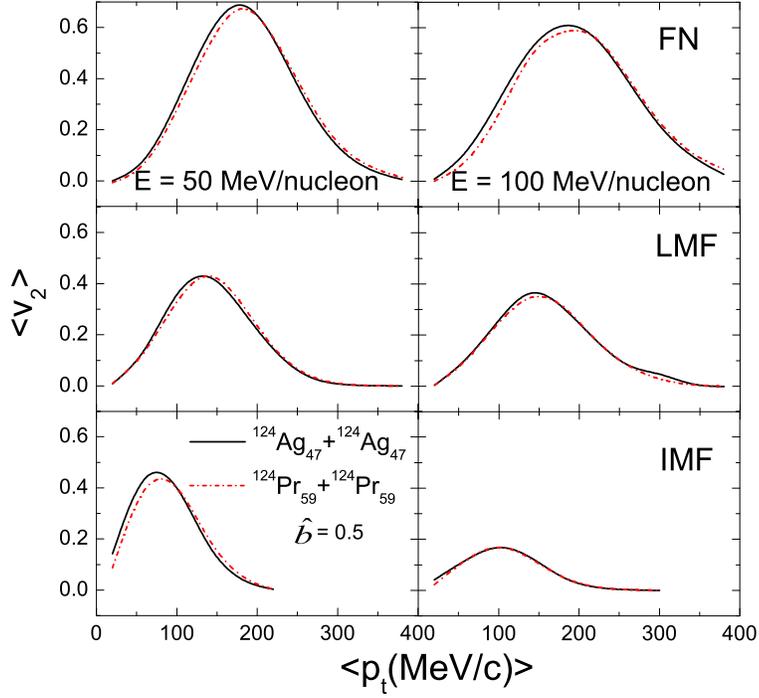}
\caption{\label{Fig:1} Transverse momentum dependence of the elliptical flow, summed over the entire rapidity distribution, at $\hat{b}$ = 0.5 for three different reactions at 50 (left) and 100 (right) MeV/nucleon.}
\end{figure}

\begin{figure}
\hspace{-2.0cm}\includegraphics[scale=0.45]{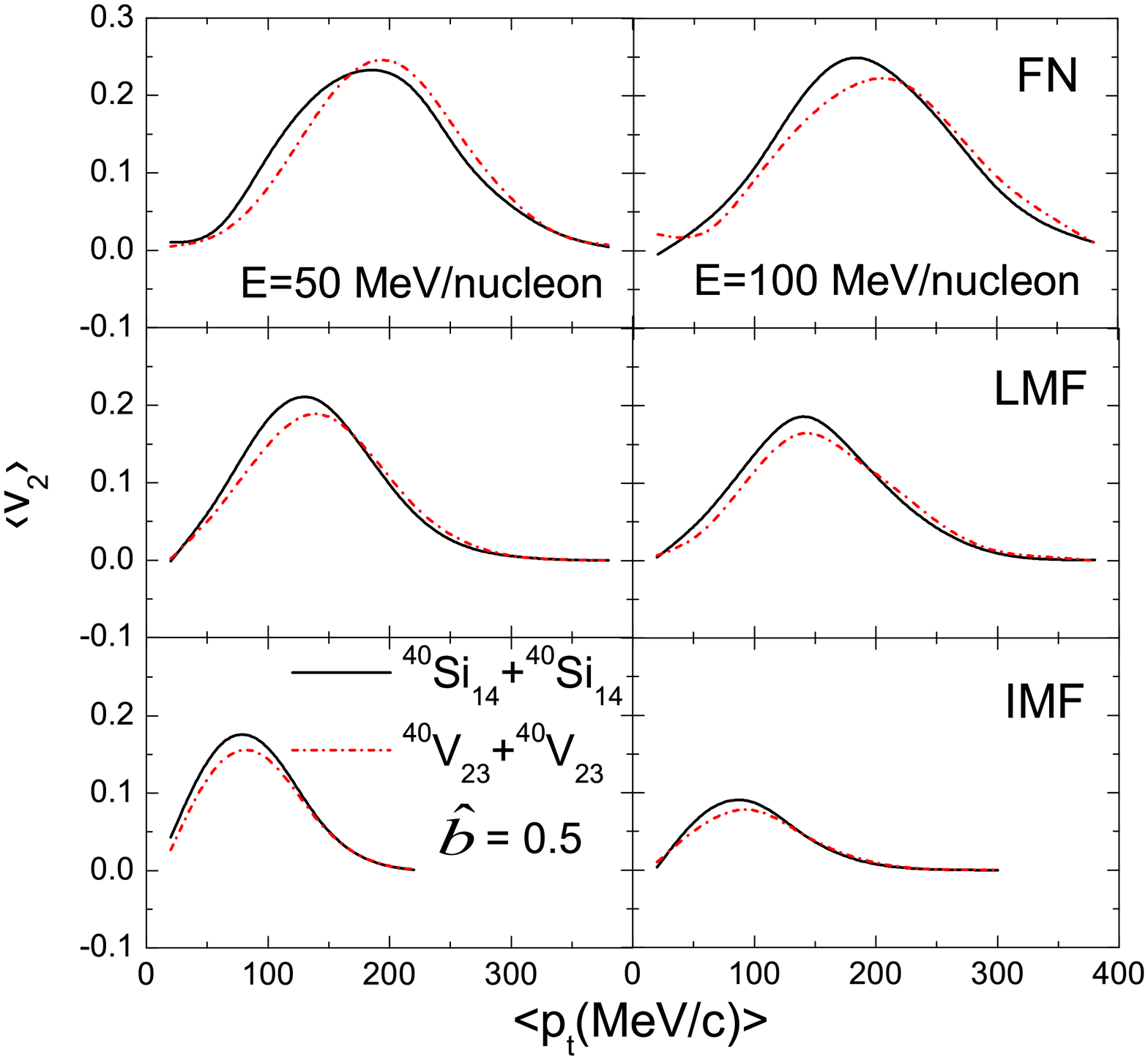}
\caption{\label{Fig:2} Transverse momentum dependence of the elliptical flow, summed over the entire rapidity distribution, at $\hat{b}$ = 0.5 for three different reactions at 50 (left) and 100 (right) MeV/nucleon.}
\end{figure}

To study the effect of charge asymmetry on the elliptical flow as a function of $\langle$ $P_{t}$ $\rangle$, (where $\langle$ $P_{t}$ $\rangle$ is transverse momentum of particle and is given by $p_{t}=\sqrt{(p_x^2 + p_y^2)}$). We display, final state elliptical flow in Fig.1 and 2 for free particles (A = 1) (upper panel), LMF's (2 $\leq$ A $\leq$ 4)(middle) and IMF's (5 $\leq$ A $\leq$ $A_{tot}$/6) (lower panel) at an incident energy E = 50 MeV/nucleon (left) and E = 100 MeV/nucleon (right) for the reactions of $^{124}X_{m}+^{124}X_{m}$, where m = 47 and 59 (in Fig.1) and $^{40}Y_{n}+^{40}Y_{n}$, where n = 14 and 23 (in Fig.2). Here elliptical flow is summed over all rapidity bins. Figs.1 and 2 reveal:\\
(a) Gaussian shape is obtained for $\langle$ $v_{2}$ $\rangle$ in all cases. Note that the elliptical flow is integrated over the entire rapidity range. This Gaussian shaped behavior is quite similar to the one reported by Colona and Di Toro et al., \cite{11}.\\
(b) Peak of the Gaussian shifts towards lower value of $\langle$ $P_{t}$ $\rangle$ for the heavier fragments. This is because the free nucleons and LMF's feel the mean field directly, while heavy fragments have weaker sensitivity \cite{12}.\\
(c) The neutron rich systems $^{124}Ag_{47}+^{124}Ag_{47}$ and $^{40}Si_{14}+^{40}Si_{14}$ exhibit weaker squeeze-out flow compared to the neutron deficient reactions $^{124}Pr_{59}+^{124}Pr_{59}$ and $^{40}V_{23}+^{40}V_{23}$. Our findings are in agreement with Zhang et al., \cite{13} where a neutron rich system was found to exhibits weaker squeeze-out flow.
%%%%%%%%%%%%%%%%%%%%%%%%%%%%%%%%%%%%%%%%%%%%%%%%%%%%%%%%%%%%%%%%%%%%%%%
 \begin{figure}
\hspace{-2.0cm}\includegraphics[scale=0.45]{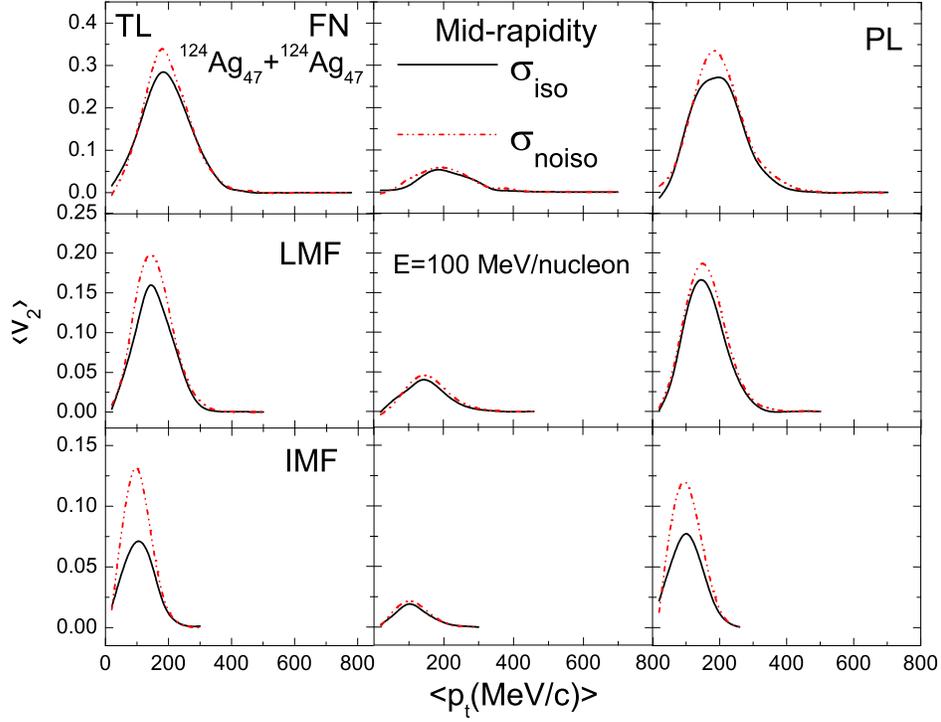}
\caption{\label{Fig:3} Transverse momentum dependence of the elliptical flow at E = 100 MeV/nucleon for two different cross-sections for the reaction $^{124}Ag_{47}+^{124}Ag_{47}$.}
\end{figure}

\begin{figure}
\hspace{-2.0cm}\includegraphics[scale=0.45]{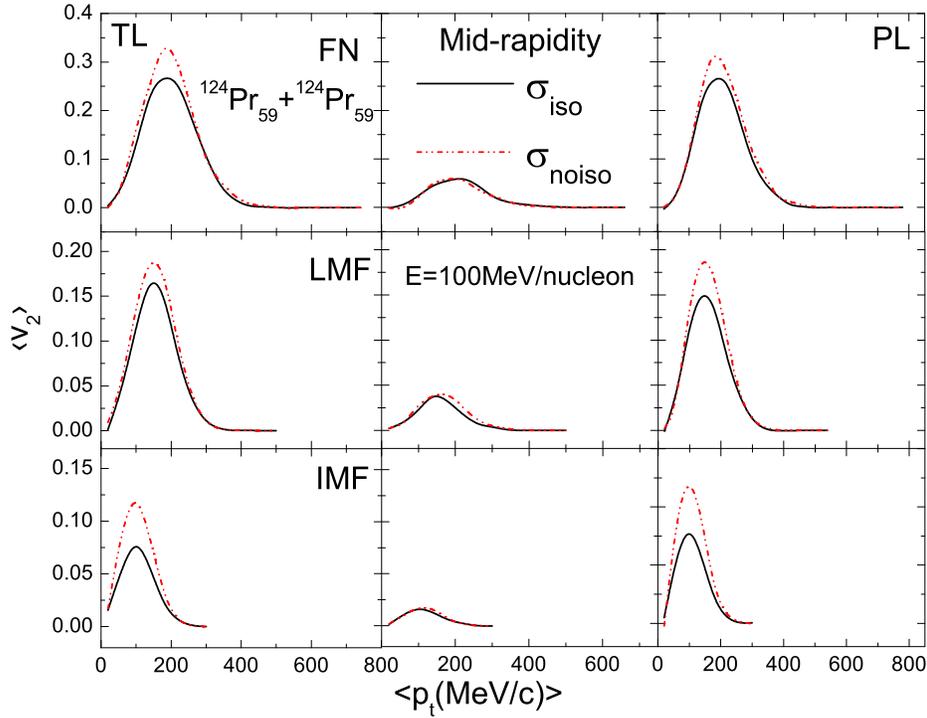}
\caption{\label{Fig:4} Same as Fig.3 but for the reaction $^{124}Pr_{59}+^{124}Pr_{59}$.}
\end{figure}

\begin{figure}
\hspace{-2.0cm}\includegraphics[scale=0.45]{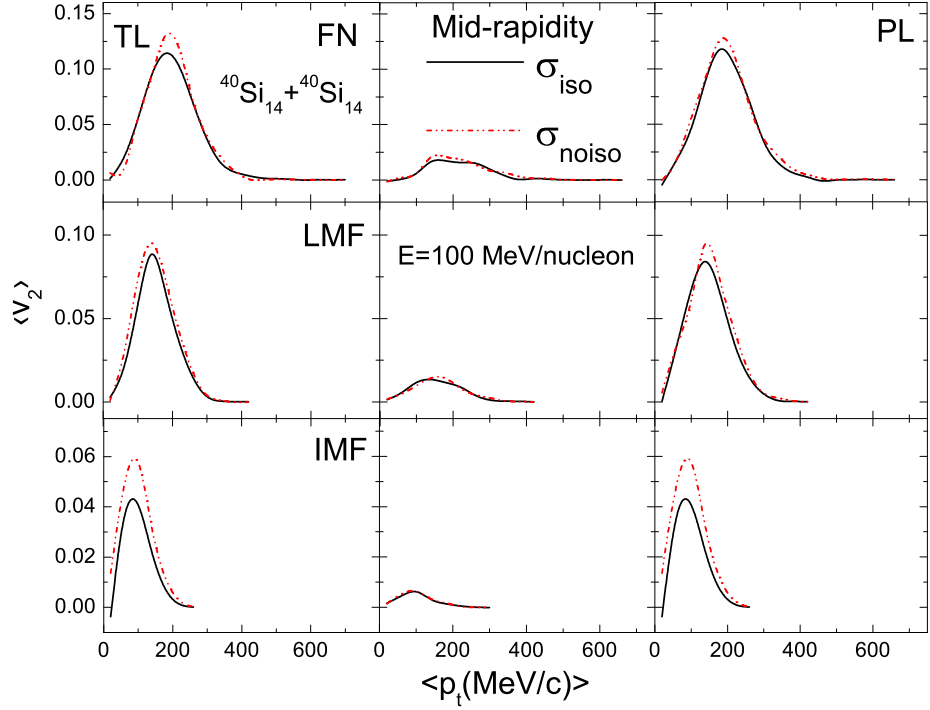}
\caption{\label{Fig:5} Same as Fig.3 but for the reaction $^{40}Si_{14}+^{40}Si_{14}$.}
\end{figure}

\begin{figure}
\hspace{-2.0cm}\includegraphics[scale=0.45]{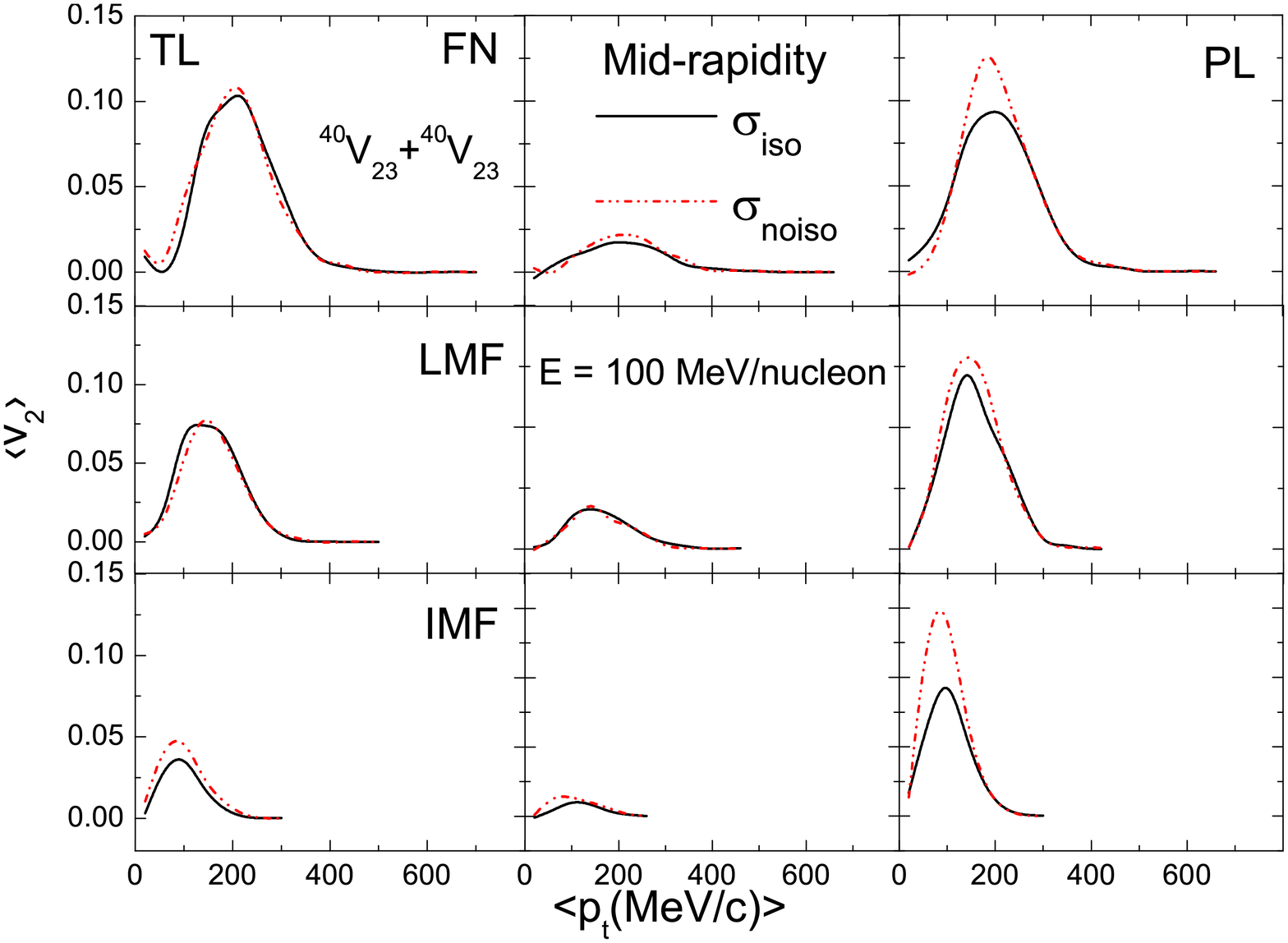}
\caption{\label{Fig:6} Same as Fig.3 but for the reaction $^{40}V_{23}+^{40}V_{23}$.}
\end{figure}

To study the effect of isospin dependence of cross-section and charge asymmetry on the elliptical flow, we display in Figs.3-6, the transverse momentum dependence of elliptical flow for the reactions of $^{124}Ag_{47}+^{124}Ag_{47}$, $^{124}Pr_{59}+^{124}Pr_{59}$, $^{40}Si_{14}+^{40}Si_{14}$ and $^{40}V_{23}+^{40}V_{23}$. We divided total elliptical flow into contributions from target-like (TL) (left panels), mid-rapidity (middle panels) and projectile-like (PL) (right panels) particles at E = 100 MeV/nucleon. The upper, middle and bottom panels represent the free nucleons, LMF's and IMF's. These figures reveal following points:\\
(a) $\langle$ $v_{2}$ $\rangle$ is sensitive to different nucleon-nucleon cross-sections. Weaker squeeze-out flow is observed in the case of isospin independent cross-section. This happens because in the case of isospin dependent cross-section, neutron-proton cross-section is three times larger compared to neutron-neutron and proton-proton cross-section that will enhance binary collisions. These findings are in agreement with ref. \cite{14}.\\
(b)  Comparison of Figs. 3 $\&$ 4 and Figs. 5 $\&$ 6 indicates significant dependence of $\langle$ $v_{2}$ $\rangle$ on the charge asymmetry of colliding pairs. Our findings are also supported by Zhang et al., \cite{13}.\\
(c) Closer looks to these figures help us to understand the origin of these isospin effects. From the figures, it is clear that percentage change is maximum in the case of mid-rapidity region for both isospin dependent and isospin independent cross-section. Indicating that isospin effects originates from mid-rapidity region.\\

%%%%%%%%%%%%%%%%%%%%%%%%%%%%%%%%%%%%%%%%%%%%%%%%%%%%%%%%%%%%%%%%

\begin{figure}
\hspace{-2.0cm}\includegraphics[scale=0.45]{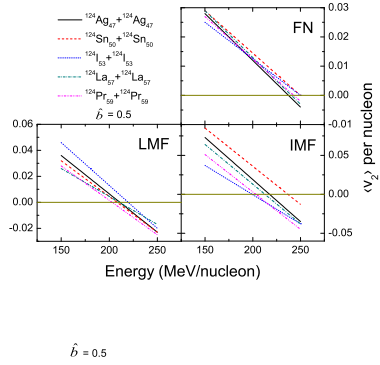}
\caption{\label{Fig:7} Variation of elliptical flow with energy for the reactions $^{124}X_{m}+^{124}X_{m}$, where $^{124}X_{m}$ = ($^{124}Ag_{47}$, $^{124}Sn_{50}$, $^{124}I_{53}$, $^{124}La_{57}$ and $^{124}Pr_{59}$).}
\end{figure}

\begin{figure}
\hspace{-2.0cm}\includegraphics[scale=0.45]{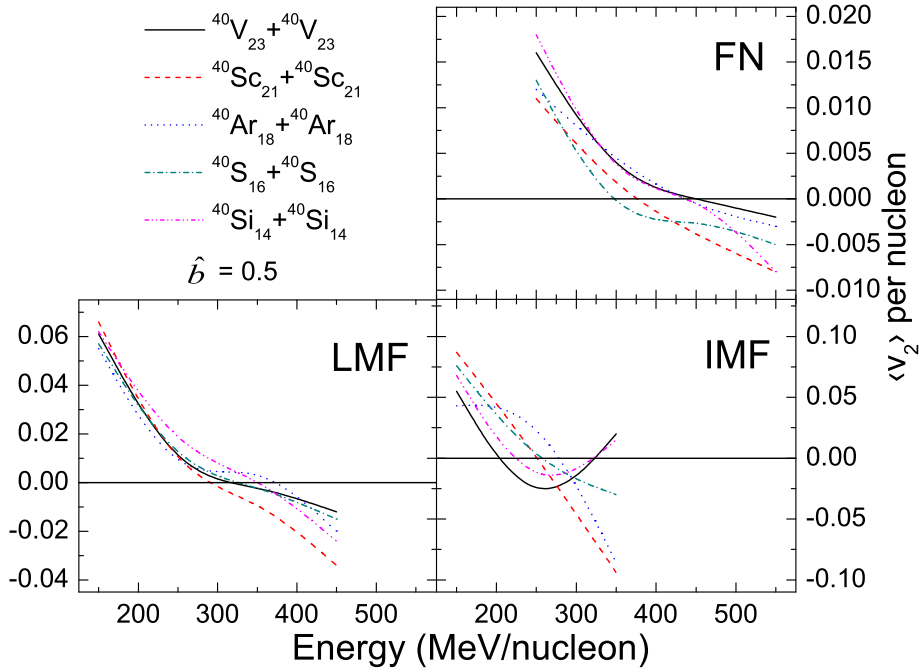}
\caption{\label{Fig:8} Variation of elliptical flow with energy for the reactions $^{40}Y_{n}+^{40}Y_{n}$, where $^{40}Y_{n}$ = ($^{40}V_{23}$, $^{40}Sc_{21}$, $^{40}Ar_{18}$, $^{40}S_{16}$ and $^{40}Si_{14}$).}
\end{figure}

In figs. 7 and 8, we display the variation of excitation function of elliptical flow $\langle$ $v_{2}$ $\rangle$ for free nucleons, LMF's and IMF's for mid-rapidity region using same set of the reactions considered earlier. We note:\\
(a) The elliptical flow turns negative with beam energy. This is because spectators move faster after $\langle$ $v_{2}$ $\rangle$ reaches a minimum value \cite{15}. The energy at which this behavior changes is found to decrease with the size of the fragment. This means that the flow of heavier fragments is larger than that of LMF's and free nucleons at all beam energies. These findings are in agreement with ref. \cite{16}.\\
(b) There occurs a transition from in-plane emission to out-of-plane. This is due to the fact that the contribution of participant zone dominates the reaction in midrapidity region leading to the transition from in-plane to out-of-plane. In other words, participant zone is primarily responsible for the transition from in-plane to out-of-plane. The energy at which this transition observed is dubbed as transition energy $E_{Trans}$.\\    
 
%%%%%%%%%%%%%%%%%%%%%%%%%%%%%%%%%%%%%%%%%%%%%%%%%%%%%%%%%%%%%%%%%%%%%%%%% 
\begin{figure}
\hspace{-2.0cm}\includegraphics[scale=0.45]{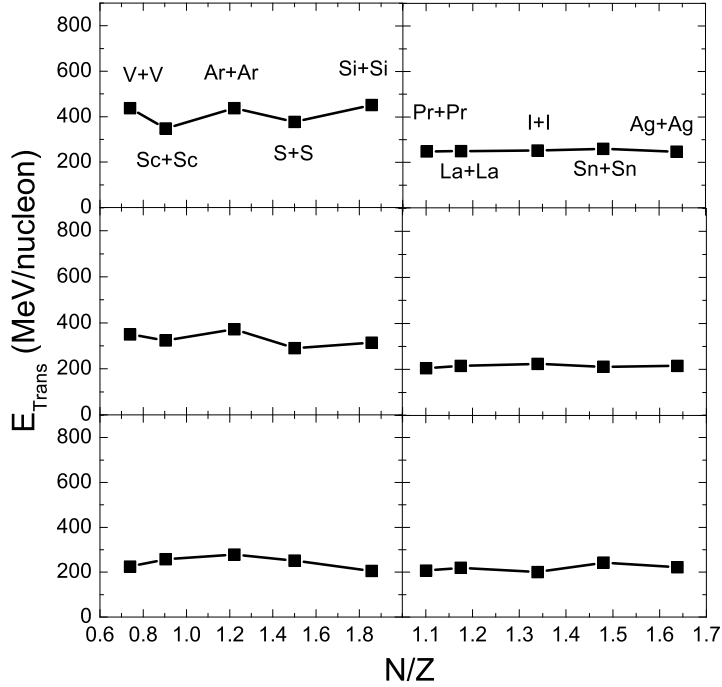}
\caption{\label{Fig:9} Transition energies ($E_{Trans}$ in MeV/nucleon) for elliptical flow as a function of N/Z.} 
\end{figure}

Now to study the effect of charge asymmetry on the transition energy of free nucleons, LMF's and IMF's, we show in Fig.9, the transition energy as a function of N/Z for two sets of reaction. From both the figures, it is clear that transition energy remains almost constant with increase in the N/Z of the system.\\

%%%%%%%%%%%%%%%%%%%%%%%%%%%%%%%%%%%%%%%%%%%%%%%%%%%%%%%%%%%%%%%%%%%%%%%%%
\begin{figure}
\hspace{-2.0cm}\includegraphics[scale=0.45]{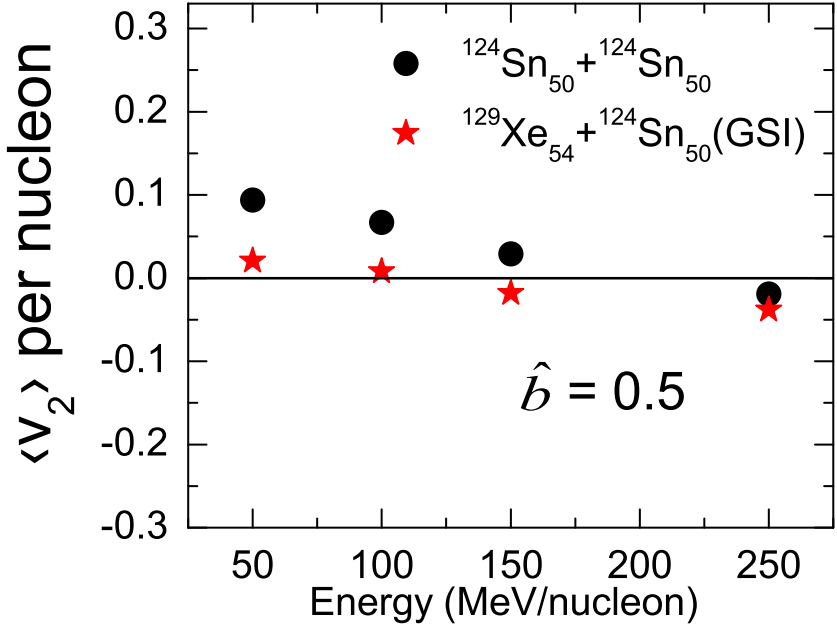}
\caption{\label{Fig:10} The energy dependence of elliptical flow for the reaction $^{124}Sn_{50}+^{124}Sn_{50}$.}
\end{figure}

To further strengthen our interpretation of the results of elliptical flow $\langle$ $v_{2}$ $\rangle$, we display in Fig.10 a comparison of theoretical results of elliptical flow with experimental data extracted by the INDRA$@$(GSI+GANIL) collaboration \cite{19}. Here simulations are performed for the reaction $^{124}Sn_{50}+^{124}Sn_{50}$ with  $\sigma_{iso}$ reduced by 20$\%$. It is worth mentioning that the results with above choice of cross-section are in good agreement with the experimental data of ref. \cite{17}. The choice of reduced cross-section has also been motivated by ref. \cite{18} as well as many previous studies \cite{19}.\\

%%%%%%%%%%%%%%%%%%%%%%%%%%%%%%%%%%%%%%%%%%%%%%%%%%%%%%%%%%%%%%%%%%%%%%%%%
%%%%%%%%%%%%%%%%%%%%%%%%%%%%%%%%%%%%%%%%%%%%%%%%%%%%%%%%%%%%%%%%%%%%%%%%%
\section{Summary}

Using the isospin dependent quantum molecular dynamics model, we have studied the effect of charge asymmetry and isospin dependent cross-section on different aspects of elliptical flow. Here simulations have been carried out for $^{124}X_{m}+^{124}X_{m}$, where m = (47, 50, 53, 57 and 59) and $^{40}X_{n}+^{40}X_{n}$, where n= (14, 16, 18, 21 and 23) . Our study shows that elliptical flow is observed to be strongly dependent on the isospin-dependent cross-section. The transition energy remains almost constant with increase in N/Z of the system. The comparison with experimental data of INDRA$@$(GSI+GANIL) collaboration supports our findings. Moreover, our results of $^{40}X_{m}+^{40}X_{m}$ will be of great use for the experimentalists working at SCC500 at VECC Kolkata (INDIA).\\

%%%%%%%%%%%%%%%%%%%%%%%%%%%%%%%%%%%%%%%%%%%%%%%%%%%%%%%%%%%%%%%%%%%%%%%%%%
{\large{\bf Acknowledgment}}

This work has been supported by a grant from the university grant commission, Government of India [Grant No. 39-858/2010(SR)].\\

\noindent

%%%%%%%%%%%%%%%%%%%%%%%%%%%%%%%%%%%%%%%%%%%%%%%%%%%%%%%%%%%%%%%%%%%%%%%%%%

\end{document}